\documentclass[preprint,nofootinbib,prd,aps]{revtex4}
\usepackage{graphicx,amstext,amssymb,amsmath,color}
\begin{document}

\title{Fixed particle number constraint in a simple  model of a thermal expanding system      and 
 $pp$ collisions at the LHC  }

\author{M.D. Adzhymambetov$^{1}$}
\author{S.V. Akkelin$^{1}$}
\author{Yu.M. Sinyukov$^{1}$${^,}$${^2}$}
\affiliation{$^1$Bogolyubov Institute for Theoretical Physics,
Metrolohichna  14b, 03143 Kyiv,  Ukraine,\\
$^2$ ExtreMe Matter Institute EMMI,
GSI Helmholtzzentrum fuer Schwerionenforschung,
Planckstrasse 1,
64291 Darmstadt,
Germany}

\begin{abstract}

Two-boson momentum correlations at fixed particle 
number constraint are  studied in a simple analytically solvable model  of  a thermal expanding system. 
We show  that the increase of expansion rate, as well as increase of particle  multiplicity,
enhances  the ground-state contribution  to particle
momentum spectra and  leads 
to suppression of the Bose-Einstein momentum correlations.   The relations of these findings
to the  multiplicity-dependent measurements
of the  Bose-Einstein momentum correlations in high-multiplicity  $p+p$ collision events  at the LHC are discussed.

\end{abstract}

\pacs{}

 \maketitle

\section{Introduction}

 Notwithstanding the evidence of the hydrodynamic expansion  in high-multiplicity $p+p$ collision events
 at the CERN Large Hadron Collider (LHC), for  recent reviews see, e.g.,  Refs. \cite{Hydro-pp-1,Hydro-pp-2},  
 robust interpretation of the multiplicity-dependent Bose-Einstein momentum correlations of 
 identical particles  created in such collisions  is still absent. To a good extent this is due to the fact that
 some peculiarities of  the  data, such as  
saturation effect in the multiplicity dependence of the
interferometry correlation radius parameters (so-called HBT radii)  for  high
charged-particle multiplicity  \cite{Atlas,CMS},  as well as low values 
of the correlation strength parameter $\lambda$ \cite{Atlas,CMS},  are at variance
with the expected behavior for emission from hydrodynamically expanding thermalized  systems.\footnote{Let us 
make a reservation, however, that a semiquantitative prediction of the  saturation effect for the 
interferometric radii at large multiplicities in $p + p$ collisions, assuming thermalization and 
hydrodynamic expansion of the arising system, was undertaken, in fact, in Ref. \cite{PBM}.}
A systematic and quantitative analysis and theoretical interpretation of the multiplicity-dependent 
Bose-Einstein momentum correlations  may therefore  elucidate the nature of the particle emitter 
in $p+p$ collisions at the LHC in crucial ways.

In a recent paper \cite{Akkelin-1} fixed particle number constraint was  applied  to a quantum-field
thermal  state of the   nonrelativistic ideal gas
of bosons at fixed temperature trapped by means of a
harmonic chemical potential.   It was demonstrated in this paper that increase with  $N$ of the particle
number density  is accompanied for fairly high $N$  by the noticeable ground-state  Bose-Einstein condensation,
and that such a condensation leads to
suppression  of the two-boson momentum correlation function. It is worth noting that this effect 
 takes place at fixed $N$ in the thermal ensemble where averaged over all multiplicities  
mean particle number density is below the critical one and, therefore, there is no grand-canonical 
ground-state  condensation.
 In the present work  we  further develop 
 the model of Ref. \cite{Akkelin-1}  aiming to account for the 
 system's expansion and thereby to bring the  model closer
to $p+p$ collision experiments. We find an exact analytical solution of the quantum thermal model with the 
system's expansion 
and show that suppression of the Bose-Einstein momentum correlations,   i.e., decrease of the $\lambda$ parameter,
is increased  if intensity of the flow  increases. We attribute  such a
suppression to  the  ground-state   contribution to particle momentum spectra  
and suggest that certain
features of the  multiplicity-dependent  two-boson  momentum correlations at high multiplicities in $p+p$ 
collisions at the LHC can be interpreted  as a signature of the  presence of  ground-state condensate.

\section{Quasiequilibrium state   of expanding nonrelativistic boson field }

We begin with a brief overview of a standard procedure for constructing  a relevant statistical 
operator $\rho$ (see, e.g., Ref. \cite{Zubarev}). 
It has been   known for a long time (see Ref. \cite{Jaynes})
that for a given set of relevant observables $A_{n}$, the actual expectation values 
of which $\langle A_{n} \rangle (t)$ are known
at some given time, the statistical operator $\rho(t)$, ``least biased'' as 
for unmonitored degrees of freedom, can be
 found from the maximum of the von Neumann   
entropy, $S=-Tr [\rho \ln{\rho}]$, subject to  the constraints  
\begin{eqnarray}
\langle A_{n} \rangle (t)= Tr [A_{n} \rho(t)], \label{1.0} \\
Tr [ \rho (t)]=1.
\label{1}
\end{eqnarray}
 This can be done  by varying the functional $S'[\rho']$, where $a_{n}(t)$ and $(\Phi (t)- 1)$ are Lagrange multipliers,
\begin{eqnarray}
S'[\rho']= - Tr [ \rho'\ln{\rho'}]-\sum_{n}a_{n}(Tr[A_{n} \rho']-\langle A_{n} \rangle)-(\Phi - 1)(Tr[\rho']-1),
\label{1.1}
\end{eqnarray}
with respect to $\rho'$ 
and then  putting variation $\delta S'[\rho']$ equal to zero, 
\begin{eqnarray}
\delta S'[\rho']= - Tr \left[ \left (\ln{\rho'} + \Phi + \sum_{n}a_{n}A_{n} \right ) \delta \rho' \right ] =0.
\label{1.2}
\end{eqnarray}
It yields  
\begin{eqnarray}
\rho(t)=\frac{1}{Z}\exp \left ( - \sum_{n} a_{n} (t) A_{n} \right ),
\label{2.1} 
\end{eqnarray}
where 
\begin{eqnarray}
Z= \exp{(\Phi)}=Tr\left[\exp \left ( - \sum_{n} a_{n} (t) A_{n} \right )\right]
\label{2.2}
\end{eqnarray}
is the normalizing  factor making  $Tr[\rho(t)]=1$. 
The statistical operator (\ref{2.1}) is sometimes called the relevant statistical operator.
One can see that it corresponds to the generalized Gibbs  state described 
by some set of observables.  If a true state of a system is unknown or very complicated, then one can utilize 
reduced (incomplete) description (\ref{2.1}), characterized by the
knowledge of mean values  of some observables only,  to make a reasonable estimate of any 
other observable $B$ at that time by means of the equation $\langle B \rangle (t)= Tr [B \rho(t)]$.
The crucial point here is the choice of the set of relevant observables, which is adequate for the reduced description
of a system. For example, for a thermalized hydrodynamically expanding system, the relevant
observables are the energy-momentum tensor and currents of conserved quantities. 
The corresponding relevant statistical operator for such a 
system is called sometimes the ``quasiequilibrium statistical operator''.

Here, to make the problem tractable, we choose as relevant observables mean values of energy 
and momentum density, as well as  particle number density,  of a free nonrelativistic scalar field defined 
at some moment of time.  We begin with the Lagrangian density 
 for a free real relativistic scalar field in Minkowski spacetime $x^{\mu}=(t,\textbf{r})$, $\textbf{r}=(x,y,z)$
 (we use the convention $g^{\mu \nu}=\mbox{diag}(+1,-1,-1,-1)$), 
\begin{eqnarray}
L=\frac{1}{2} \left (\frac{\partial
\phi}{\partial t}\right )^{2} - \frac{1}{2} \left (\frac{\partial
\phi}{\partial \textbf{r}}\right )^{2} - \frac{m^{2}}{2}\phi^{2}.
\label{3}
\end{eqnarray}
The canonical momentum field is then $\pi = \dot \phi$, where an
overdot denotes a derivative with respect to time, and the  Hamiltonian density is given by 
\begin{eqnarray}
H=\frac{1}{2}\pi^{2} + \frac{1}{2} ({\bf \nabla } \phi ) ^{2} + \frac{m^{2}}{2}\phi^{2},
\label{4}
\end{eqnarray}
where $\mathbf{\nabla}=(\partial_{1},\partial_{2},\partial_{3})=(d/dx,d/dy,d/dz)$.
The corresponding energy-momentum tensor reads
\begin{eqnarray}
T^{\mu \nu}(x)=\partial^{\mu}\phi\partial^{\nu}\phi -g^{\mu \nu}L.
\label{5}
\end{eqnarray}

There are different methods to arrive
at an effective nonrelativistic description for a real scalar field. 
Typically these methods start with an
appropriate field redefinition.  Here, for the sake of convenience, we relate the complex nonrelativistic 
field $\Psi$ to the real relativistic field $\phi$ by using the relations (see, e.g., Ref. \cite{Guth-2})
\begin{eqnarray}
\phi (t, {\bf r} )= \frac{1}{\sqrt{2m} } \left ( e^{-imt} \Psi (t, {\bf r}) + e^{imt} \Psi^{\dag}  (t, {\bf r} ) \right), \label{6} \\
\pi (t, {\bf r} ) = - i \sqrt{ \frac{ m}{2} }  \left ( e^{-imt} \Psi (t, {\bf r}) - e^{imt} \Psi^{\dag}  (t, {\bf r} ) \right ). \label{7}
\end{eqnarray}
Equations  (\ref{6}) and (\ref{7}) 
give a one-to-one mapping between the   complex-valued $\Psi$ and the real-valued $\phi$ and its conjugate
momentum $\pi$. The quantization prescriptions
$[\phi (t, {\bf r} ),\pi (t, {\bf r}' )]=i\delta^{(3)}({\bf r}-{\bf r}')$,  
$[\phi (t, {\bf r} ),\phi (t, {\bf r}' )]=[\pi (t, {\bf r} ),\pi (t, {\bf r}' )]=0$ result in 
the commutation relations
\begin{eqnarray}
[\Psi (t,\textbf{r}), \Psi^{\dag} (t, \textbf{r}') ] =
\delta^{(3)}(\textbf{r} - \textbf{r}' ), \label{7.1}
\end{eqnarray}
and
\begin{eqnarray}
[\Psi (t,\textbf{r}), \Psi (t,\textbf{r}')  ] =[\Psi^{\dag}
(t,\textbf{r}), \Psi^{\dag} (t,\textbf{r}')  ]= 0 . \label{7.2}
\end{eqnarray}
The Fourier-transformed operators are defined as
\begin{eqnarray}
\Psi (t, \textbf{p}) = (2\pi)^{-3/2}\int d^{3}r
e^{-i\textbf{p}\textbf{r}} \Psi ( t,\textbf{r}), \label{7.3} \\
\Psi^{\dag}  (t, \textbf{p}) = (2\pi)^{-3/2}\int d^{3}r
e^{i\textbf{p}\textbf{r}} \Psi^{\dag} (t, \textbf{r}). \label{7.4}
\end{eqnarray}
They satisfy the following canonical commutation relations:
\begin{eqnarray}
[\Psi (t, \textbf{p}), \Psi^{\dag} (t, \textbf{p}') ] =
\delta^{(3)}(\textbf{p} - \textbf{p}' ),  \label{7.5}
\end{eqnarray}
and
\begin{eqnarray}
 [\Psi (t, \textbf{p}), \Psi (t, \textbf{p}') ] =[\Psi^{\dag} (t, \textbf{p}), \Psi^{\dag} (t, \textbf{p}') ]= 0.  \label{7.6}
\end{eqnarray}

In order to take the nonrelativistic limit in the Hamiltonian density,  one can  substitute Eqs.
(\ref{6}) and (\ref{7})  into Eq. (\ref{4}). The corresponding expression contains rapidly oscillating 
terms proportional to $e^{ \pm  2imt}$. These terms are usually neglected in the nonrelativistic limit,
because in the limit of large $m$ they average to zero on timescales larger than $1/m$, and 
the remaining terms are expected to be slowly varying
compared to the timescale $1/m$. We then obtain the following nonrelativistic Hamiltonian density\footnote{
It is noteworthy that this expression can be also obtained as a leading term in a low-energy
($p^{2}/m^{2} \ll 1$) expansion by using a  nonlocal field redefinition proposed 
in Ref. \cite{Guth-2} instead of the local one defined  in Eqs. (\ref{6}) and (\ref{7}).}: 
\begin{eqnarray}
H(t,\textbf{r})=T^{00}(t,\textbf{r})=\frac{1}{2m}{\bf \nabla}  \Psi {\bf \nabla } \Psi^{\dag} + m \Psi \Psi^{\dag}.
\label{8}
\end{eqnarray}
 In a similar way,  using Eq. (\ref{5}) and neglecting terms with fast oscillatory factors $e^{ \pm  2imt}$, we obtain  
that ($j=1,2,3$)
\begin{eqnarray}
T^{0j}(t,\textbf{r})=-\frac{i}{2}\left ( \Psi \partial^{j} \Psi^{\dag} - \Psi^{\dag} \partial^{j} \Psi  \right ).
\label{9}
\end{eqnarray}
In the nonrelativistic approximation, the particle number density is given by 
\begin{eqnarray}
N (t,\textbf{r})=\Psi  \Psi^{\dag}.
\label{10}
\end{eqnarray}

The quasiequilibrium statistical operator, associated with the expectation values of  the
selected observables (\ref{8})-(\ref{10}), reads\footnote{For simplicity, we will suppress the dependence on $t$.}
\begin{eqnarray}
\rho =\frac{1}{Z}\hat{\rho},
\label{11}
\end{eqnarray}
where  
$Z$ is the  partition function,
\begin{eqnarray}
Z = Tr[\hat{\rho}], \label{11.1}
\end{eqnarray}
and
\begin{eqnarray}
\hat{\rho}=\exp \left [ -  \int d^{3}r \beta (\textbf{r})\left (\frac{1}{\sqrt{1-\textbf{u}^{2}(\textbf{r})}} T^{00}+ \frac{u_{j}(\textbf{r})}{\sqrt{1-\textbf{u}^{2}(\textbf{r})}}  T^{0j} -\mu (\textbf{r})\Psi  \Psi^{\dag} \right ) \right ].
\label{11.2}
\end{eqnarray}
Here $\beta =1/T$ is inverse temperature, ${\bf u}=(u^{1},u^{2},u^{3})$ ($u_{j}=-u^{j}$)  is collective velocity,
$\mu$ is chemical potential, $j=1,2,3$, and summation of repeated indices is implied. 

Given (\ref{8}) and  (\ref{9}), we may now consider the effective
description for such a model in the nonrelativistic limit when 
$\frac{1}{\sqrt{1-\textbf{u}^{2}}}\approx (1+\frac{1}{2}\textbf{u}^{2})$
 and
$\frac{\textbf{u}}{\sqrt{1-\textbf{u}^{2}}}\approx \textbf{u}$.
 Then, taking into account that  $\textbf{u}^{2}(\frac{1}{2m}{\bf \nabla}  \Psi {\bf \nabla } \Psi^{\dag}+ m \Psi \Psi^{\dag})\approx \textbf{u}^{2}m \Psi \Psi^{\dag}$ in the  nonrelativistic limit of  large $m$,   we obtain 
 \begin{eqnarray}
\left(1+\frac{1}{2}\textbf{u}^{2}(\textbf{r})\right)T^{00}\approx \frac{1}{2m}{\bf \nabla}  \Psi {\bf \nabla } \Psi^{\dag} + 
 m (1+\frac{1}{2}\textbf{u}^{2}(\textbf{r}))\Psi \Psi^{\dag}.
\label{11.3}
\end{eqnarray}
   Finally, taking into account Eqs. (\ref{8}), (\ref{9}), and (\ref{11.3}), one can rewrite (\ref{11.2}) in the form 
\begin{eqnarray}
\hat{\rho}=\exp \left [ -  \int d^{3}r \beta (\textbf{r}) \left ( \frac{1}{2m}[-i{\bf \nabla }-m{\bf u} (\textbf{r}) ]\Psi[i {\bf \nabla}-m{\bf u}(\textbf{r}) ]\Psi^{\dag}- \hat{\mu} (\textbf{r})\Psi \Psi^{\dag}    \right ) \right ],
\label{12}
\end{eqnarray}
where 
\begin{eqnarray}
\hat{\mu}= \mu - m .
\label{13}
\end{eqnarray}

If actual  expectation values $\langle T^{0j} \rangle  =0$, then  ${\bf u} = {\bf 0}$ and the 
quasiequilibrium statistical operator 
(\ref{12}) describes   a nonexpanding system of 
noninteracting nonrelativistic  bosons. 
Below we demonstrate that the  quasiequilibrium statistical operator of an expanding system, see Eq.
 (\ref{12}), can be rewritten  to such a form by means of a 
simple unitary  redefinition of $\Psi$ and $\Psi^{\dag}$ fields, if the collective velocity ${\bf u}$ belongs to 
the class of the potential velocity fields,
\begin{eqnarray}
\textbf{u} (\textbf{r}) = -\frac{1}{m}{\bf \nabla } \theta (\textbf{r}) ,
\label{13.1}
\end{eqnarray}
where $\theta$ is a dimensionless flow potential. For this aim  let  us 
rewrite the operator-valued fields  $\Psi$ and 
$\Psi^{\dag}$ in terms of the fields $\hat{\Psi}$ and 
$\hat{\Psi}^{\dag}$ as follows: 
\begin{eqnarray}
\Psi ({\bf r} ) = e^{-i\theta (\bf{r})} \hat{\Psi}(\bf{r}) , \label{14} \\
\Psi^{\dag} ({\bf r} ) = e^{i\theta (\bf{r})} \hat{\Psi}^{\dag}(\bf{r}). \label{15}
\end{eqnarray}
One can see that 
\begin{eqnarray}
[\hat{\Psi} (\textbf{r}), \hat{\Psi}^{\dag} ( \textbf{r}') ] =
\delta^{(3)}(\textbf{r} - \textbf{r}' ), \label{15.1}
\end{eqnarray}
and
\begin{eqnarray}
[\hat{\Psi} (\textbf{r}), \hat{\Psi} (\textbf{r}')  ] =[\hat{\Psi}^{\dag}
(\textbf{r}), \hat{\Psi}^{\dag} (\textbf{r}')  ]= 0 . \label{15.2}
\end{eqnarray}
Substituting (\ref{14}) and (\ref{15}) into (\ref{12}) 
and accounting for Eq. (\ref{13.1}), we have
\begin{eqnarray}
\hat{\rho}=\exp \left [ -  \int d^{3}r \beta (\textbf{r})\left ( \frac{1}{2m}[-i{\bf \nabla }]\hat{\Psi}[i {\bf \nabla}]\hat{\Psi}^{\dag}- \hat{\mu} (\textbf{r})\hat { \Psi} \hat{\Psi}^{\dag}    \right ) \right ].
\label{17}
\end{eqnarray}
Hence,   we obtain that the quasiequilibrium statistical operator $\rho=\hat{\rho}/Z$   describes the expanding state of
field $\Psi$, see Eq. (\ref{12}), and   
nonexpanding state  of the transformed field
$\hat{\Psi}$, see Eq.  (\ref{17}). 

\section{ Fixed particle number constraint in  exactly solvable 
model of the quasiequilibrium state}

Calculations of expectation values with statistical operator are significantly 
simplified if the corresponding statistical operator  can be diagonalized in some 
representation. To make it possible with the quasiequilibrium statistical operator $\rho$ , see Eqs. (\ref{11}),
(\ref{12}), and (\ref{17}), below we assume that $\beta =1/T$ is constant 
and that the  chemical potential $\hat{\mu} (\textbf{r})$  reads
\begin{eqnarray}
\hat{\mu} (\textbf{r})= -
\frac{m}{2}(\omega_{x}^{2}x^{2}+\omega_{y}^{2}y^{2}+\omega_{z}^{2}z^{2}) + \mu^{*},
\label{18}
\end{eqnarray}
where $\mu^{*}=\mbox{const}$. Such a choice for the chemical potential means a ``harmonic trap'' distribution 
of particles.  Then Eq. (\ref{17}) takes the form 
\begin{eqnarray}
\hat{\rho} = e^{- \beta K }, \label{19} 
\end{eqnarray}
where $K$ is defined by
\begin{eqnarray}
K = \int d^{3}r \hat{\Psi}^{\dag}
(\textbf{r})\left(-\frac{1}{2m}\nabla^{2} - \hat{\mu} (\textbf{r})\right
)\hat{\Psi}(\textbf{r}). \label{20}
\end{eqnarray}
It is worth noting that $K$ does not commute with the Hamiltonian $H=\int d^{3}r H(t,\textbf{r})$,
see Eq. (\ref{8}).  Therefore,  the zero-temperature ground state of the statistical operator  is
an eigenstate of $K$ but not of $H$.

It is well known that $K$, see Eqs.  (\ref{18}) and (\ref{20}), can be diagonalized in the
oscillator representation
\begin{eqnarray}
\hat{\Psi}(\textbf{r})=
\sum_{n,k,l=0}^{\infty}\alpha(n,k,l)\phi_{n}(x)\phi_{k}(y)\phi_{l}(z),
\label{20.1}
\end{eqnarray}
where the creation $\alpha^{\dag}(n,k,l)$ and annihilation
$\alpha(n,k,l)$ operators satisfy the commutation relations
\begin{eqnarray}
[\alpha(n,k,l),\alpha^{\dag}(n',k',l')] = \delta_{nn'}
\delta_{kk'}\delta_{ll'} , \label{20.2}
\end{eqnarray}
and
\begin{eqnarray}
 [\alpha(n,k,l),\alpha(n',k',l')] =[\alpha^{\dag}(n,k,l),\alpha^{\dag}(n',k',l')]= 0 . \label{20.3}
\end{eqnarray}
Functions   $\phi_{n}(x)$, $\phi_{k}(y)$, and $\phi_{l}(z)$ are the
harmonic oscillator  eigenfunctions, for example, 
\begin{eqnarray}
 \phi_{n}(x) =(2^{n}n!\pi^{1/2}b_{x})^{-1/2}H_{n}
 \left(\frac{x}{b_{x}}\right)\exp\left( -\frac{1}{2}\left( \frac{x}{b_{x}}\right)^{2}\right), \label{20.4}
\end{eqnarray}
where $H_{n}(x/b_{x})$ is the Hermite polynomial, and
\begin{eqnarray}
\epsilon_{n}=\omega_{x}\left(n+\frac{1}{2}\right), \label{20.5} \\
b_{x}=(m\omega_{x})^{-1/2}. \label{20.6}
\end{eqnarray}
In such a basis, the $K$ reads
\begin{eqnarray}
K =
\sum_{n,k,l=0}^{\infty}(\epsilon_{n}+\epsilon_{k}+\epsilon_{l}-\mu^{*})\alpha^{\dag}(n,k,l)
\alpha(n,k,l). \label{20.7}
\end{eqnarray}
This implies that statistical operator $\rho=\hat{\rho}/Z$  
involves states with a various number of particles $N$ and describes a grand-canonical ensemble.\footnote{Note
that the  mean
particle number $\langle N \rangle$, defined by the  grand-canonical ensemble, as well as the 
particle number $N$,  are the same for 
$\Psi$,  $\hat{\Psi}$,   and 
$\alpha$ particles  because  transformations (\ref{14}) and (\ref{20.1})
do not mix creation and annihilation operators and preserve the standard commutation relations.} To consider 
the canonical subensemble, where the number of particles 
is fixed, one needs 
to make the corresponding projection and  define
the canonical statistical operator,
\begin{eqnarray}
\rho_{N} = \frac{1}{Z_{N}} \hat{\rho}_{N} ,  \label{21}
\end{eqnarray}
which corresponds to
subensemble of events with fixed particle number constraint. 
Here  $\hat{\rho}_{N} = {\cal P}_{N} \hat{\rho }{\cal P}_{N}$, 
where ${\cal P}_{N}$ is the projection operator that  automatically invokes the corresponding constraint, and 
$Z_{N} $ is the corresponding  canonical partition function that is needed to insure the probability interpretation of
the ensemble obtained in the result of this projection, $Z_{N}=Tr[\hat{\rho}_{N}]$. Corresponding formalism was
developed in Ref. \cite{Akkelin-1} 
and the reader is referred to this paper
for details of the calculations.
Taking into account  that $\rho_{N}$ does not depend on $\mu^{*}$ (this dependence is factored out in Eq. (\ref{21})),
one can rewrite 
Eq. (\ref{21})  as 
\begin{eqnarray}
\rho_{N} = \frac{1}{Z_{N}^{0}} \hat{\rho}_{N}^{0} .  \label{21.1}
\end{eqnarray}
Here we denote $\hat{\rho}_{N}$ and $Z_{N}$ associated with $\mu^{*}=0$ as $\hat{\rho}_{N}^{0}$ and $Z_{N}^{0}$,
respectively. The  canonical 
partition functions satisfy  the recursive formula  \cite{Recurr-1}
\begin{eqnarray}
n Z_{n}^{0}= \sum_{s=1}^{n} \sum_{\textbf{j}} e^{-s\beta \epsilon_{\textbf{j}}}Z_{n-s}^{0},
\label{21.2}
\end{eqnarray}
where  $\epsilon_{\textbf{j}}=\epsilon_{n,k,l}=\epsilon_{n}+\epsilon_{k} +
\epsilon_{l}$, $Z_{0}^{0}=\langle 0 | 0 \rangle =1$, and $n=1,...,N$.

The goal of this study is to evaluate   the two-boson momentum correlation  functions at 
fixed multiplicities for  an expanding system. Such a  correlation function is defined as the ratio of
the two-particle momentum spectrum to one-particle ones and can be
written   at fixed multiplicities as
\begin{eqnarray}
C_{N}(\textbf{k},\textbf{q})=G_{N} \frac{\langle
\Psi^{\dag}(\textbf{p}_{1})\Psi^{\dag}(\textbf{p}_{2})\Psi(\textbf{p}_{1})\Psi(\textbf{p}_{2})\rangle_{N}}{\langle
\Psi^{\dag}(\textbf{p}_{1})\Psi(\textbf{p}_{1})\rangle_{N}\langle
\Psi^{\dag}(\textbf{p}_{2})\Psi(\textbf{p}_{2})\rangle_{N}}.
\label{22}
\end{eqnarray}
Here and below $\langle  ... \rangle_{N} \equiv Tr[\rho_{N}  ...]$,
${\bf k}=({\bf
p}_{1}+{\bf p}_{2})/2$, ${\bf q}={\bf p}_{2}-{\bf p}_{1}$, and
$G_{N}$ is the normalization constant. The latter  is needed  to
normalize the theoretical Bose-Einstein correlation function in accordance with
normalization that is applied by experimentalists:
$C^{exp}(\textbf{k},\textbf{q}) \rightarrow
 1$ for $|\textbf{q}| \rightarrow \infty$.

 The assumption of the  potential velocity field (\ref{13.1}) and transformations (\ref{14}) and  (\ref{15})
 allow one to apply for calculations of the one- and two- particle spectra the same technique
 that was used  in Ref. \cite{Akkelin-1} for a nonexpanding system. We start  by using   
expectation values  of operators $\alpha$  and $\alpha^{\dag}$ 
(these expressions  are calculated in Ref. \cite{Akkelin-1} and are provided below for the reader's convenience),
 \begin{eqnarray}
\langle
\alpha^{\dag}(\textbf{j}_{1})\alpha(\textbf{j}_{2})\rangle_{N} =
\delta_{\textbf{j}_1\textbf{j}_2} \sum_{s=1}^{N} e^{-s\beta 
\epsilon_{\textbf{j}_{2}}}\frac{Z_{N-s}^{0}}{Z_{N}^{0}}, 
\label{23} \\
\langle
\alpha^{\dag}(\textbf{j}_{1})\alpha^{\dag}(\textbf{j}_{2})\alpha(\textbf{j}_{3})\alpha(\textbf{j}_{4})\rangle_{N}
= \nonumber \\ 
(\delta_{\textbf{j}_1\textbf{j}_4}\delta_{\textbf{j}_2\textbf{j}_3}
+
\delta_{\textbf{j}_1\textbf{j}_3}\delta_{\textbf{j}_2\textbf{j}_4})\sum_{s=1}^{N-1}
\sum_{s'=1}^{N-s} e^{-s\beta \epsilon_{\textbf{j}_{4}}}e^{-s' \beta 
\epsilon_{\textbf{j}_{3}}}\frac{Z_{N-s-s'}^{0}}{Z_{N}^{0}},
\label{24}
\end{eqnarray}
where for notational simplicity we write $\textbf{j}$ instead of $(n,k,l)$.
The next step is to utilize Eqs. (\ref{23}), (\ref{24}),  and (\ref{20.1}) to obtain expectation values 
$\langle \hat{\Psi}^{\dag}({\bf{r}}_{1}) \hat{\Psi}({\bf{r}}_{2}) \rangle_{N}$  and 
$\langle \hat{\Psi}^{\dag}({\bf{r}}_{1}) \hat{\Psi}^{\dag}({\bf{r}}_{2}) \hat{\Psi}({\bf{r}}_{3}) \hat{\Psi}({\bf{r}}_{4}) \rangle_{N}$. Then, for example,
\begin{eqnarray}
\langle \hat{\Psi}^{\dag}(\textbf{r}_{1}) \hat{\Psi} (\textbf{r}_{2})\rangle_{N}
= \nonumber \\ \sum_{s=1}^{N}\frac{Z_{N-s}^{0}}{Z_{N}^{0}}
\sum_{n=0}^{\infty}\sum_{k=0}^{\infty}\sum_{l=0}^{\infty}\phi_{n}^{*}(x_{1})\phi_{k}^{*}(y_{1})\phi_{l}^{*}(z_{1})
\phi_{n}(x_{2})\phi_{k}(y_{2})\phi_{l}(z_{2})e^{-\frac{3}{2}s\beta
\omega}e^{-s \beta \omega (n+k+l)}, \label{24.1}
\end{eqnarray}
where the eigenfunctions  are defined by Eq. (\ref{20.4}).
In order to keep things as simple as possible, here and below we assume
that $\omega_{x}=\omega_{y}=\omega_{z}=\omega$, then $b_{x}=b_{y}=b_{z}=b$ and
\begin{eqnarray}
b = (m\omega)^{-1/2},
\label{27.1}
\end{eqnarray}
see Eqs. (\ref{20.4}) and  (\ref{20.6}). Utilizing the integral representation of the 
Hermite function,
\begin{eqnarray}
H_{n}\left( \frac{x}{b}\right )= \left ( \frac{b}{i}\right
)^{n}\frac{b}{2\sqrt{\pi}}e^{\frac{x^{2}}{b^{2}}}
\int_{-\infty}^{+\infty}v^{n}e^{-\frac{1}{4}b^{2}v^{2}+ixv}dv ,
\label{27.1-1}
\end{eqnarray} 
one can simplify corresponding 
expressions. This was done in Ref. \cite{Akkelin-1}, and the results can be written as  
\begin{eqnarray}
\langle \hat{\Psi}^{\dag}(\textbf{r}_{1}) \hat{\Psi} (\textbf{r}_{2})\rangle_{N}
= \sum_{s=1}^{N} \frac{Z_{N-s}^{0}}{Z_{N}^{0}} \hat{\Phi} (\textbf{r}_{1},\textbf{r}_{2},\beta \omega s),  \label{27.2-1} \end{eqnarray}
and
\begin{eqnarray}
\langle \hat{\Psi}^{\dag}(\textbf{r}_{1})\hat{\Psi}^{\dag}(\textbf{r}_{2})\hat{\Psi}(\textbf{r}_{3})\hat{\Psi}(\textbf{r}_{4})\rangle_{N}=
\nonumber
\\
\sum_{s=1}^{N-1}\sum_{s'=1}^{N-s}\frac{Z_{N-s-s'}^{0}}{Z_{N}^{0}}\left( \hat{\Phi} (\textbf{r}_{1},\textbf{r}_{3},\beta \omega s)
\hat{\Phi} (\textbf{r}_{2},\textbf{r}_{4},\beta \omega s') + \hat{\Phi} (\textbf{r}_{1},\textbf{r}_{4},\beta \omega s)
\hat{\Phi} (\textbf{r}_{2},\textbf{r}_{3},\beta \omega s')\right), \label{27.2-2} 
\end{eqnarray}
where
\begin{eqnarray}
\hat{\Phi} (\textbf{r}_{1},\textbf{r}_{2},\beta \omega s)= \frac{1}{(2\pi)^{3/2}}\frac{1}{b^{3}}\left
(\sinh(\beta \omega s)\right )^{-3/2}\exp\left ( -
\frac{\textbf{r}_{1}^{2}+\textbf{r}_{2}^{2}}{2b^{2}\tanh(\beta
\omega s) } \right )\exp\left(
\frac{\textbf{r}_{1}\textbf{r}_{2}}{b^{2}\sinh(\beta \omega
s)}\right ). \label{27.2-3}
\end{eqnarray}

To  evaluate  particle momentum spectra for an
expanding system,  one needs first  to specify
 the velocity  profile. Here, for the purpose of illustration, we chose flow profile in the  linear isotropic form,
\begin{eqnarray}
    {\bf u} ({\bf r}) = \kappa {\bf r}.
 \label{27.3}   
\end{eqnarray}
Then the solution of Eq. (\ref{13.1}) can be written as 
\begin{eqnarray}
\theta ({\bf r}) = -\frac{m\kappa{\bf r}^2}{2} .
\label{27.4}
\end{eqnarray}
Equation (\ref{27.4}) allows us  to relate $\Psi$ with $\hat{\Psi}$, see Eqs. (\ref{14}) and  (\ref{15}),
and thereby  to  define the expectation values
$\langle \Psi^{\dag}(\textbf{r}_{1}) \Psi(\textbf{r}_{2}) \rangle_{N}$  and 
$\langle \Psi^{\dag}(\textbf{r}_{1}) \Psi^{\dag}(\textbf{r}_{2}) \Psi(\textbf{r}_{3}) \Psi(\textbf{r}_{4}) \rangle_{N}$  
for an expanding system. 
It is then a simple matter to perform Fourier transformations and 
calculate one- and two-particle momentum spectra which are defined as corresponding expectation values
for Fourier-transformed 
field operators (\ref{7.3}) and (\ref{7.4}).  The results  are  
\begin{eqnarray}
\langle \Psi^{\dag}(\textbf{p}_{1}) \Psi (\textbf{p}_{1})\rangle_{N}
=
\sum_{s=1}^{N}\frac{Z_{N-s}^{0}}{Z_{N}^{0}}\Phi(\textbf{p}_{1},\textbf{p}_{1},\beta
\omega s, \kappa),
 \label{28} 
\end{eqnarray}
and
\begin{eqnarray}
\langle
\Psi^{\dag}(\textbf{p}_{1})\Psi^{\dag}(\textbf{p}_{2})\Psi(\textbf{p}_{1})\Psi(\textbf{p}_{2})\rangle_{N}=\nonumber
\\
\sum_{s=1}^{N-1}\sum_{s'=1}^{N-s}\frac{Z_{N-s-s'}^{0}}{Z_{N}^{0}}\left (
\Phi(\textbf{p}_{1},  \textbf{p}_{1
},\beta \omega s, \kappa)\Phi(\textbf{p}_{2},
\textbf{p}_{2},\beta \omega s',\kappa) + \Phi(\textbf{p}_{1},
\textbf{p}_{2},\beta \omega s, \kappa)\Phi(\textbf{p}_{2},  \textbf{p}_{1},\beta \omega
s', \kappa)\right ), \label{29}
\end{eqnarray}
respectively. Here we introduce notation
\begin{eqnarray}
\Phi(\textbf{p}_{1},\textbf{p}_{2},\beta \omega s, \kappa) =
\frac{b^{3}(1+m^{2}\kappa^{2}b^{4})^{-3/2}}{(2\pi\sinh(\beta \omega s))^{3/2}} \times \nonumber \\ \exp\left ( -
\frac{b^{2}}{4(1+m^{2}\kappa^{2}b^{4})}\left[(\textbf{p}_{1}+\textbf{p}_{2})^{2}\tanh(\frac{\beta \omega s}{2} ) +
\frac{(\textbf{p}_{2}-\textbf{p}_{1})^{2}}{\tanh(\frac{\beta
\omega s}{2})} + 2 i (\textbf{p}_{2}^{2}-\textbf{p}_{1}^{2})m\kappa b^{2} \right  ]\right ).
 \label{30}
\end{eqnarray}
One can see that, for a nonexpanding system, i.e., for $\kappa=0$, Eqs. (\ref{28})-(\ref{30})
are reduced to the corresponding expressions presented in Ref. \cite{Akkelin-1}.\footnote{Note that here  we
slightly simplified and optimized the notations 
as compared to Ref. \cite{Akkelin-1}.} Furthermore, one can easily see that   
\begin{eqnarray}
|\Phi(\textbf{p}_{1},\textbf{p}_{2},\beta \omega s, \kappa) |= \xi^{3}|\Phi(\xi\textbf{p}_{1},\xi\textbf{p}_{2},\beta \omega s, 0)|,
 \label{30.1}
\end{eqnarray}
where
\begin{eqnarray}
\xi= \frac{1}{\sqrt{1+m^{2}\kappa^{2}b^{4}}}\leq 1.
 \label{30.2}
\end{eqnarray}

Finally, to completely  specify the two-boson correlation function (\ref{22}), one needs to estimate the  normalization
constant $G_{N}$. It can be realized by means of  the limit $|\textbf{q}| \rightarrow
\infty$ at fixed $\textbf{k}$ in the corresponding expression. One
can readily see that proper normalization is reached if 
\begin{eqnarray}
G_{N}=
\frac{Z_{N}^{0}}{Z_{N-2}^{0}}\left(\frac{Z^{0}_{N-1}}{Z^{0}_{N}}\right
)^{2}.
 \label{31}
\end{eqnarray}
This value coincides with normalization constant calculated  in Ref. \cite{Akkelin-1} for a nonexpanding 
system.
Equations (\ref{22}),  (\ref{28})-(\ref{30}), and (\ref{31}) serve in the next section 
as the starting point for the investigation of multiplicity and flow dependencies of
two-particle Bose-Einstein momentum correlations at fixed particle number constraint.

\section{Two-boson momentum correlations at fixed 
multiplicities in  the  thermal  model of an expanding system }

In the following, we focus on the multiplicity and flow dependencies of the correlation function (\ref{22}).
Below we assume that the model provides  qualitatively reasonable
estimations of these  dependencies beyond  the nonrelativistic region $p^{2}/m^{2} \ll 1$.
To discuss relations to $p+p$ collisions at the LHC,  we utilize  for numerical calculations the set of parameters
corresponding roughly to the values at the system’s breakup in $p+p$ collisions at the LHC energies.
For specificity, we take  the particle's mass as of a charged pion,  $m=139.57$ MeV, 
and the temperature $T=150$ MeV. 
Following  Ref. \cite{Akkelin-1}, we 
introduce parameter  $R$ such as   
\begin{eqnarray}
\omega= \frac{1}{R \sqrt{\beta m}}
 \label{32}
\end{eqnarray}
and treat $R$ as a free parameter instead of $\omega$. For $R$ we use $1.5$  fm.  
Using Eq. (\ref{32}), one gets 
\begin{eqnarray}
\beta \omega  = \frac{1}{R} \sqrt{\frac{\beta}{m}}=\frac{\Lambda_{T}}{R},
 \label{33}
\end{eqnarray}
and 
\begin{eqnarray}
b = \frac{1}{\sqrt{m\omega}}=\sqrt{\Lambda_{T} R},
 \label{34}
\end{eqnarray}
where $\Lambda_{T}$ is the thermal wavelength,  which we defined as
\begin{eqnarray}
\Lambda_{T} = \frac{1}{\sqrt{mT}}. 
 \label{35}
\end{eqnarray}

\begin{figure}[!ht]
\centering
\includegraphics[scale=0.7]{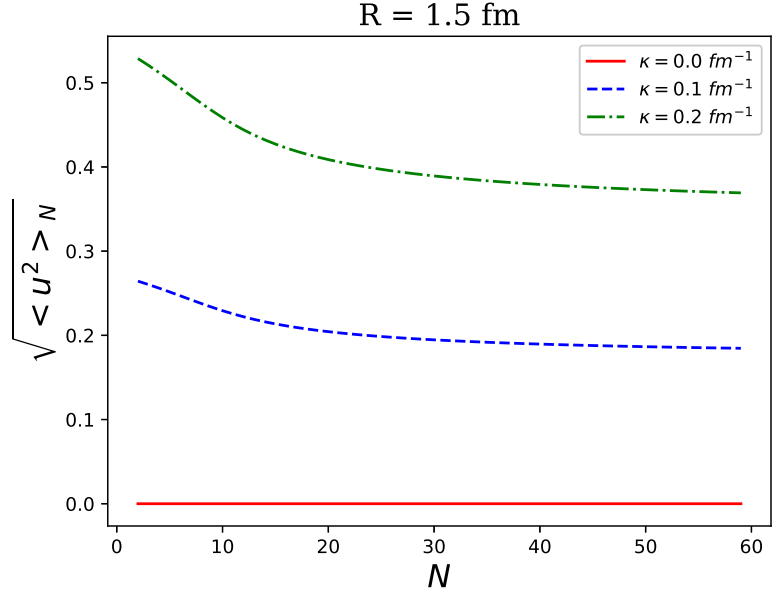}
\caption{The $\sqrt{\langle \textbf{u}^{2} \rangle_{N}}$ dependence on $N$ at different  $\kappa$. }
\label{fig:1}
\end{figure}

It is convenient to relate parameter $\kappa$ in Eq. (\ref{27.3}) with a physically meaningful
parameter in relativistic particle and nucleus collisions, namely, with mean 
flow velocity of
the system at fixed particle number constraint $\sqrt{\langle \textbf{u}^{2} \rangle}_{N}$, where
\begin{eqnarray}
 \langle \textbf{u}^{2} \rangle_{N}  = \frac{\int dxdydz
\textbf{u}^{2}\langle \Psi^{\dag}(\textbf{r}) \Psi
(\textbf{r})\rangle_{N} }{\int dxdydz \langle
\Psi^{\dag}(\textbf{r}) \Psi (\textbf{r})\rangle_{N} }. \label{36}
\end{eqnarray}
Substituting (\ref{27.3}) into the right-hand side of Eq. (\ref{36}), we have
\begin{eqnarray}
\langle \textbf{u}^{2} \rangle_{N}= \kappa^{2} \langle  \textbf{r}^{2} \rangle_{N}= 3 \kappa^{2} \langle  x^{2} \rangle_{N}, \label{37}
\end{eqnarray}
where 
\begin{eqnarray}
\langle x^{2} \rangle_{N} =\frac{\int dxdydz
x^{2}\langle \Psi^{\dag}(\textbf{r}) \Psi
(\textbf{r})\rangle_{N} }{\int dxdydz \langle
\Psi^{\dag}(\textbf{r}) \Psi (\textbf{r})\rangle_{N} }. \label{38}
\end{eqnarray}
Because of  relations (\ref{14}) and  (\ref{15}), 
$\langle
\Psi^{\dag}(\textbf{r}) \Psi (\textbf{r})\rangle_{N}  =\langle
\hat{\Psi}^{\dag}(\textbf{r}) \hat{\Psi} (\textbf{r})\rangle_{N}$, 
thereby the mean spatial  size of the system at fixed
multiplicity $\sqrt{\langle x^{2} \rangle_{N}}$ does not depend on intensity of flow. 
A corresponding expression has been calculated in Ref. \cite{Akkelin-1}. For the used set of parameter values,  
$\sqrt{\langle x^{2} \rangle_{N}}$  is close to $R$. 
Figure \ref{fig:1} shows mean flow velocity at fixed
multiplicity 
$\sqrt{\langle \textbf{u}^{2} \rangle}_{N}$ as a function of $N$ for several
different values of the strength of the
 expansion parameter $\kappa$. For  $\kappa$ we use  $0.0$,  $0.1$, 
 and $0.2$ fm$^{-1}$.
 
\begin{figure}[!ht]\
\centering
\includegraphics[scale=0.7]{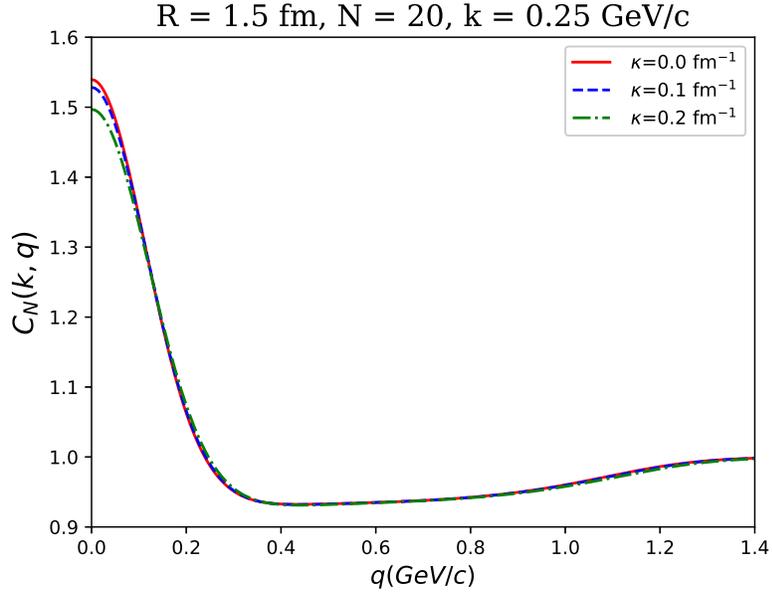}
\caption{Correlation functions  with  $k=0.25$ GeV/$c$, $N=20$,   $R=1.5$ fm at different  $\kappa$. See the text for details.} 
\label{fig:2}
\end{figure}

 Then we investigate how the two-boson momentum correlation function (\ref{22}) is affected by
 the flow. The results are plotted in Fig.  \ref{fig:2} for various values  
of $\kappa$ at $k=0.25$ GeV/$c$.  One can see that the intercept of the correlation function
$C_{N}(\textbf{k}, \textbf{0})$ decreases when the strength of the
 expansion parameter $\kappa$  increases. 
 
\begin{figure}[!ht]
\centering
\includegraphics[scale=0.7]{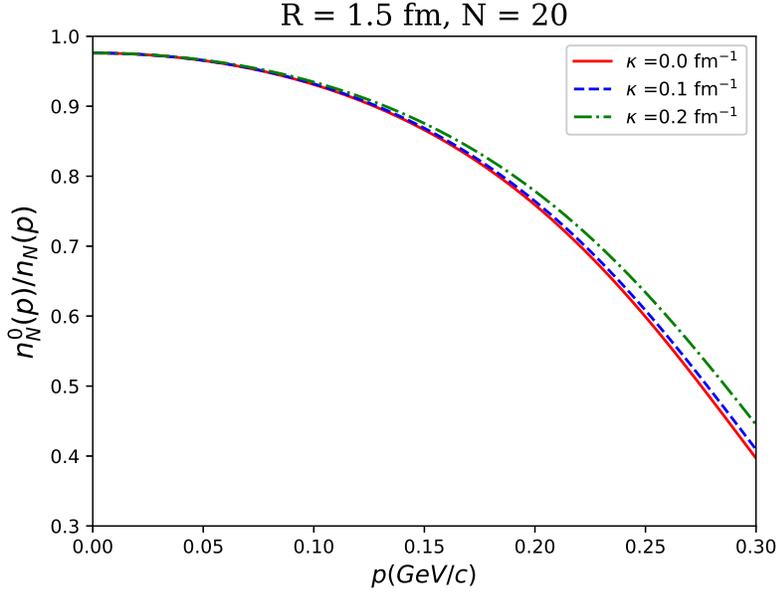}
\caption{The $n_{N}^{0}(\textbf{p})/n_{N}(\textbf{p})$ ratio as a function of particle momentum $p$ for $N=20$, 
$R=1.5$ fm, and for several different values of  $\kappa$. See the text for details.  }
\label{fig:3}
\end{figure}

To have some insight into why it happens,  it is useful to calculate the ground-state contribution 
to particle momentum spectra,
$n_{N}^{0}(\textbf{p},\kappa)$. To derive this expression, one needs to take $k=l=n=0$ in Eq. (\ref{24.1}), and 
then follow the derivation of the 
one-particle momentum spectrum
$n_{N}(\textbf{p},\kappa) = \langle \Psi^{\dag}(\textbf{p}) \Psi (\textbf{p})\rangle_{N}$, see Eq. (\ref{28}). 
The result is 
\begin{eqnarray}
    n_{N}^{0}(\textbf{p},\kappa)=\sum_{s=1}^{N} \frac{Z_{N-s}^0}{Z_{N}^{0}} \frac{b^3e^{-\frac{3}{2}\beta\omega s}}{\pi^{\frac{3}{2}}\left(1+m^2\kappa^2b^4\right)^{3/2}} \exp \left(-\frac{b^2 {\bf p}^2}{ 1 + m^2\kappa^2b^4} \right). 
     \label{40}
\end{eqnarray}
While  $n_{N}(\textbf{p},\kappa)$ and $n_{N}^{0}(\textbf{p},\kappa)$ both depend  on $\xi$ and thereby on $\kappa$,
\begin{eqnarray}
n_{N}(\textbf{p},\kappa)= \xi^{3}n_{N}(\xi\textbf{p}, 0), \label{40.1} \\
    n_{N}^{0}(\textbf{p},\kappa)= \xi^{3}n_{N}^{0}(\xi\textbf{p}, 0) ,
     \label{40.2}
\end{eqnarray}
see Eqs. (\ref{28}), (\ref{30})-(\ref{30.2}), and (\ref{40}), the occupation of the ground state
$N_{0}=\int d^{3}p n_{N}^{0}(\textbf{p},\kappa)$ and  the ground-state condensate
fraction $N_{0}/N$ do not depend on $\kappa$ at fixed $N$. On the other hand, 
because $n_{N}^{0}(\textbf{p}, 0)/n_{N}(\textbf{p},0)$ is a  decreasing function of particle 
momentum, we obtain  that $n_{N}^{0}(\xi\textbf{p}, 0)/n_{N}(\xi \textbf{p},0)$ increases when $\xi$ decreases.
Accounting for  Eq. (\ref{30.2}), we then conclude that an increase of $\kappa$ results 
in an increase of the  $n_{N}^{0}(\textbf{p},\kappa)/n_{N}(\textbf{p},\kappa)$ ratio, because  
\begin{eqnarray}
\frac{n_{N}^{0}(\textbf{p},\kappa)}{n_{N}(\textbf{p},\kappa)}= \frac{n_{N}^{0}(\xi\textbf{p}, 0)}{n_{N}(\xi\textbf{p},0)},
     \label{40.3}
\end{eqnarray}
see Eqs. (\ref{40.1}) and  (\ref{40.2}).
In Fig. \ref{fig:3} 
 we plot
 this  ratio as a function of particle momentum
 for several different values of
 the $\kappa$ parameter. The curves show that the ground-state   fraction of the particle momentum spectra
 increases at moderately high momenta 
 when  $\kappa$ increases, signaling the increasing
importance of the  ground-state contribution to particle momentum spectra. 
This implies that   particle emission at such momenta becomes more coherent when intensity of flow increases,
 leading thereby  to
 the decrease  of the intercept of the two-boson momentum correlation function: It is well known
 that the intercept of the two-boson momentum
correlation function for a chaotic emission is equal to $2$, and that the intersept 
 is equal to $1$ for a coherent emission; see, e.g., Ref. \cite{Sin-1}.

One observes from Fig.  \ref{fig:2}  the essential non-Gaussianity of the correlation functions
beyond the region of the correlation peak. Such a non-Gaussianity was discussed for a nonexpanding system 
in Ref. \cite{Akkelin-1},  
where it was demonstrated that $C_{N}(\textbf{k}, \textbf{q})$ can be rather well fitted by the two-Gaussian expression. 
If the fitting procedure is restricted to the correlation  peak region,
then the correlation function is well fitted by the one-Gaussian expression 
\begin{eqnarray}
C_{N}^{1g}(\textbf{k},\textbf{q})= 1+\lambda(\textbf{k},N)e^{-\textbf{q}^{2}R_{HBT}^{2}(\textbf{k},N)},
\label{41}
\end{eqnarray}
where  $1+\lambda(\textbf{k},N)$ is equal to the intercept of the correlation function  $C_{N}(\textbf{k},\textbf{0})$.
In order to make contact with the previous findings
of Ref. \cite{Akkelin-1}, one can relate correlation functions 
of an expanding system  $C_{N}(\textbf{k},\textbf{q},\kappa)$   with the ones for 
a nonexpanding system $C_{N}(\textbf{k},\textbf{q},0)$. This can be accomplished
using Eqs. (\ref{22}) and  (\ref{28})-(\ref{30.1}).  The result is
\begin{eqnarray}
C_{N}(\textbf{k},\textbf{q},\kappa)= C_{N}(\xi\textbf{k},\xi\textbf{q},0).
\label{42}
\end{eqnarray}
This relation means, in particular,   that 
\begin{eqnarray}
\lambda(\textbf{k},N, \kappa)= \lambda(\xi\textbf{k},N, 0), 
\label{43} \\
R_{HBT}(\textbf{k},N,\kappa) = \xi R_{HBT}(\xi \textbf{k},N,0).
\label{44}
\end{eqnarray}
It follows from Eq. (\ref{44}) that increase of $\kappa$ at fixed $\xi \textbf{k}$ results in decrease of $R_{HBT}$.   

\begin{figure}[!ht]
\centering
\includegraphics[scale=0.7]{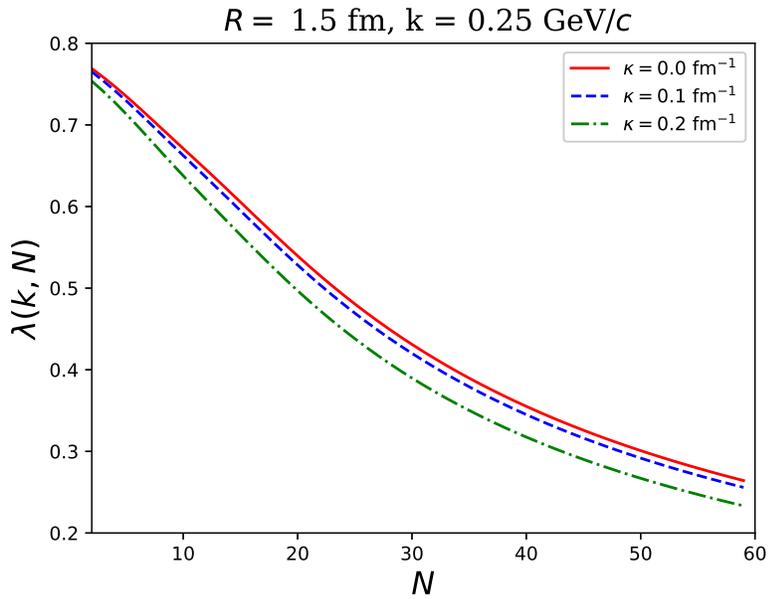}
\caption{The $\lambda$ at $ k=0.25 $ GeV/$c$ for several  different values of  $\kappa$. See the text for details. }
\label{fig:4}
\end{figure}

Figure \ref{fig:4} displays the  $\lambda$ parameter as a function of $N$ for various values
of $\kappa$. All  three curves reveal a consistent trend:  increase  of  $N$ results in decrease of
the $\lambda$ parameter, i.e., the intercept of the correlation function is reduced. Reasons
for such a behavior were discussed
in detail in  Ref. \cite{Akkelin-1}. In short, increase of $N$ results in an increase of the value 
of the ground-state fraction $N_{0}/N$,  leading for fairly high $N$ to the noticeable Bose-Einstein condensation in
the corresponding ground state of the fixed $N$ canonical ensemble state. Such
a condensation strengthens the coherent  properties of the canonical ensemble state and
results in the decrease of the intercept of the two-boson momentum correlation function
when $N$ increases.

\begin{figure}[!ht]
\centering 
\includegraphics[scale=0.5]{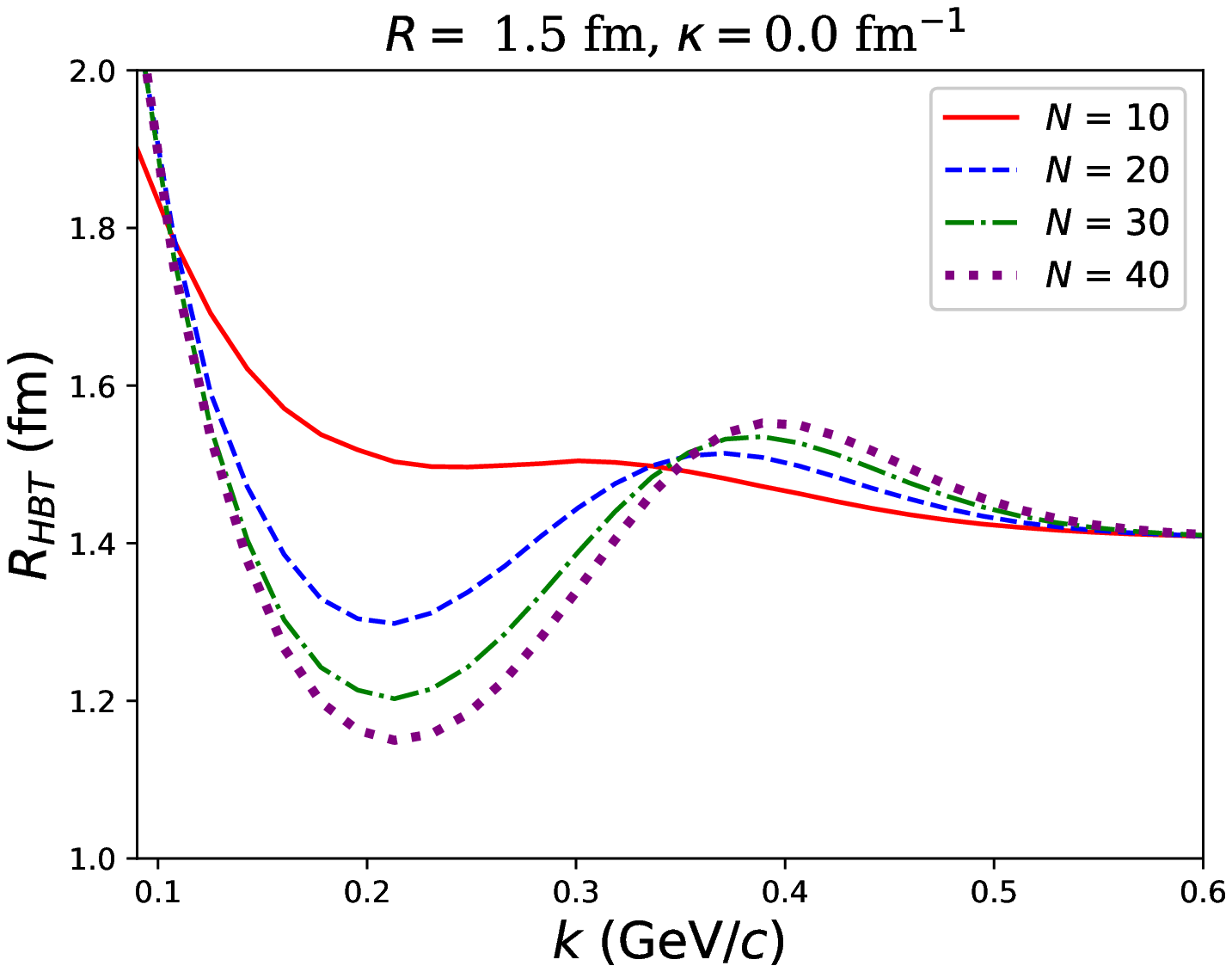} 
\includegraphics[scale=0.5]{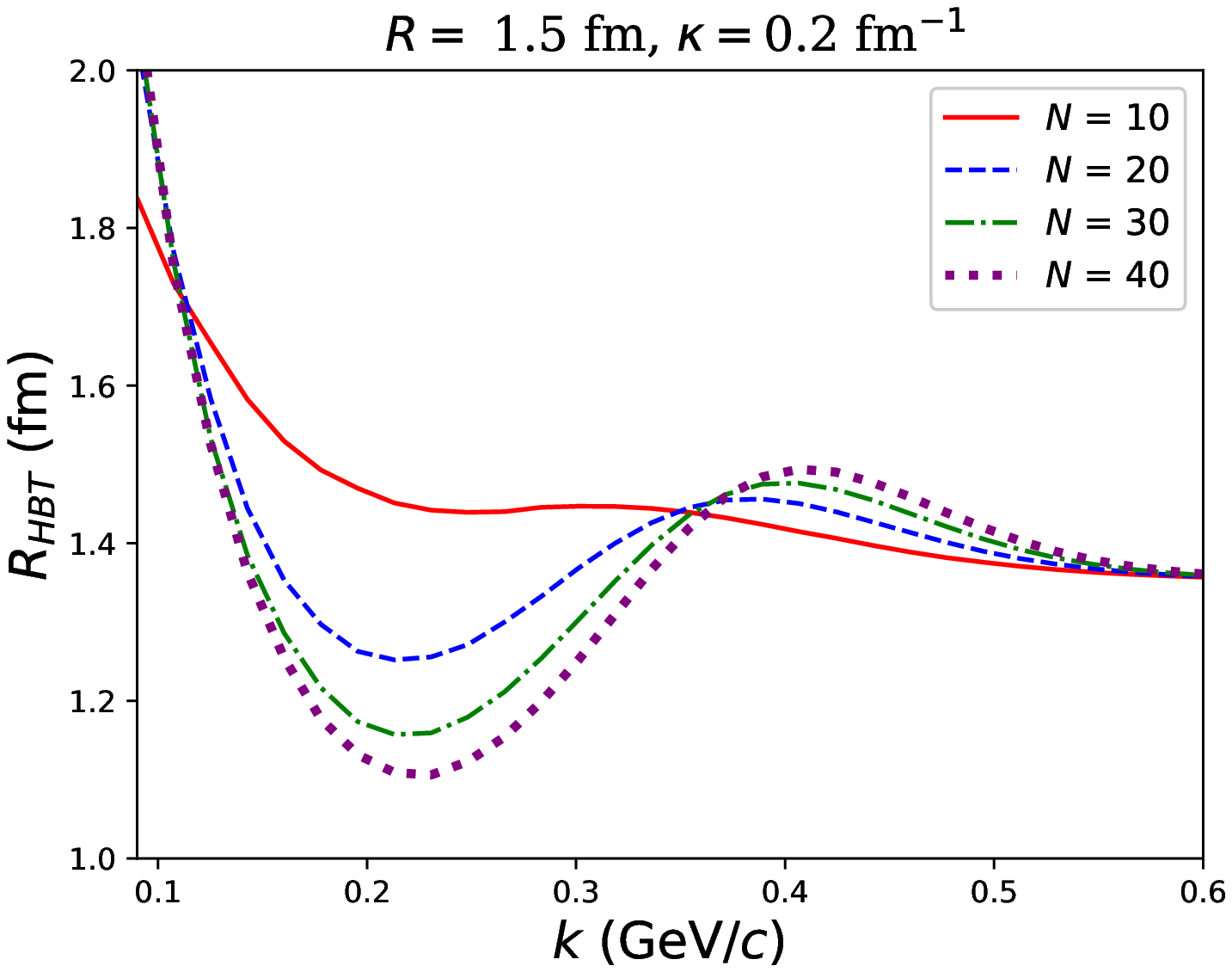}
\caption{HBT radii obtained from the one-Gaussian fit of the two-boson correlation function for several different values 
of $N$, as a function of the pair average momentum $k$.}
\label{fig:5}
\end{figure}

Figure \ref{fig:5} shows
$R_{HBT}$ for $\kappa=0.0$  and $\kappa=0.2$ fm$^{-1}$ as a function on $k$
for several different  values of $N$. One can see that, unlike  the mean spatial 
size, $R_{HBT}$ depends on the intensity of flow.\footnote{Note that decrease of interferometry radii
when intensity of flow increases   can be interpreted as
the decrease of ``homogeneity lengths'' \cite{Sin-2} (sizes of the effective emission region). }
Also, one observes from this figure  that, 
similar to  zero flow results presented in 
 Ref. \cite{Akkelin-1}, the   interferometry radii are
independent of $N$ at moderately high pair momenta.

\begin{figure}[!ht]
\centering 
\includegraphics[scale=0.5]{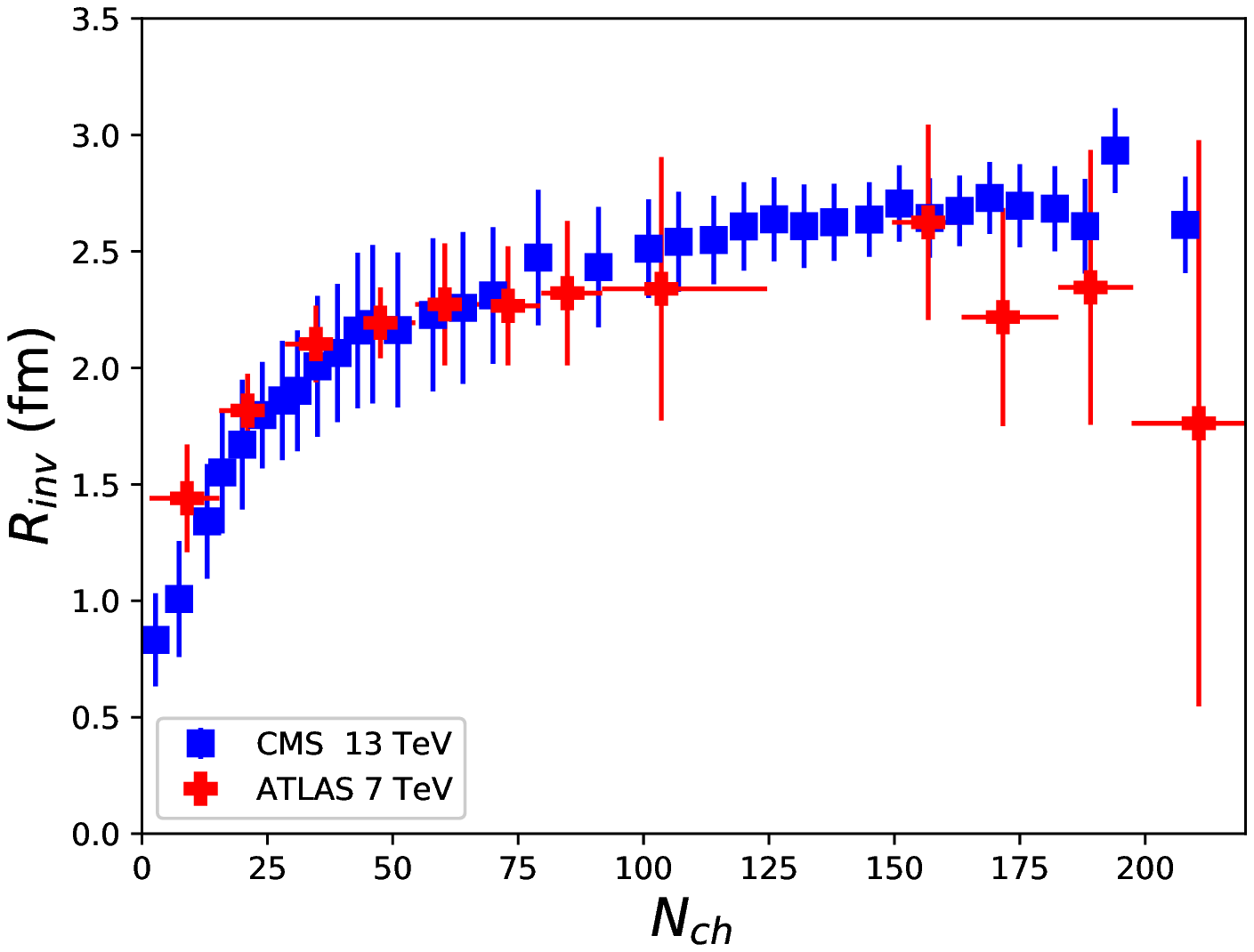} 
\includegraphics[scale=0.5]{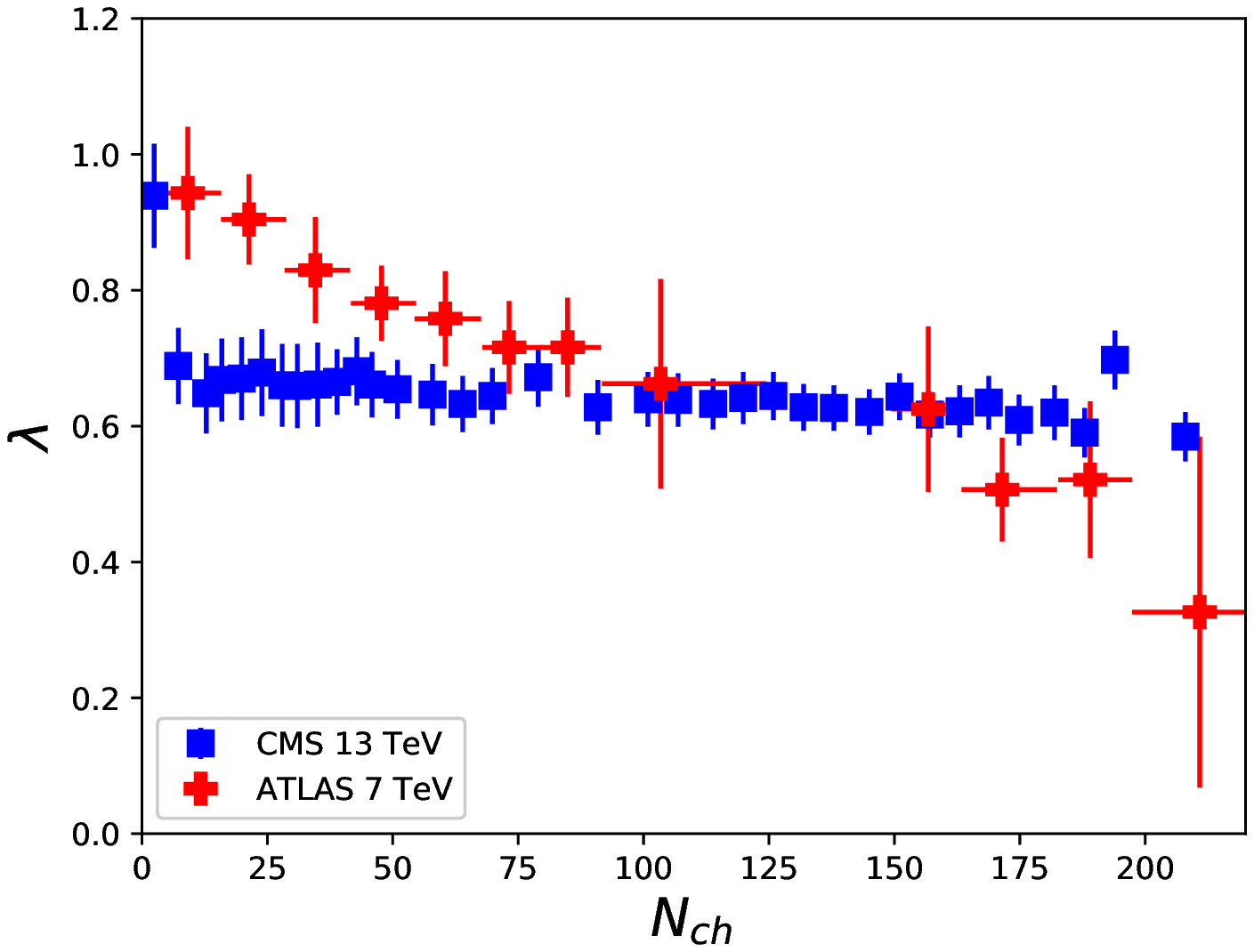}
\caption{The radius parameters obtained from exponential fits  (left) and 
$\lambda$ parameters (right), as a function of multiplicity. See Refs. \cite{Atlas,CMS} for details. }
\label{fig:6}
\end{figure}

Finally, let us discuss possible relations of this  model with high-multiplicity   $p+p$
collision events  at the LHC.  
For the purpose of illustration, we show in Fig. \ref{fig:6} some experimental data
presented by the ATLAS \cite{Atlas} and  CMS \cite{CMS} Collaborations. It is worth 
noting that analysis procedures applied by 
the ATLAS  and  CMS Collaborations are quite different, 
therefore additional adjustment (which is not done in Fig. \ref{fig:6}) 
is needed for direct comparison of the  results, see Ref. \cite{CMS}  for details. 
First of all,  one can deduce from the published data
(see Fig. \ref{fig:6}, left)  that the two-boson 
momentum correlation radius parameters  are small,\footnote{The results for exponential fits are shown. To compare
the values of the radius parameters obtained from exponential and Gaussian fits, the $R$ value of the Gaussian should
be compared with $R/\sqrt{\pi}$ of the exponential form, see Ref. \cite{Atlas}.}
compatible with the pion thermal wavelength, and 
do not change much with the collision energy. The latter seems to be natural if
the actual size of the system is related to the mean multiplicity,\footnote{We do not consider here effects conditioned 
by shape fluctuation of nucleon, see, e.g., Ref.  \cite{Hydro-pp-2} and references therein.} because at high energies, 
where increase of the collision energy
might be accompanied by the increase of the expansion  rate, the  mean multiplicity  
increases rather weakly with energy of collisions. Notice that small size of the system, together
with the  high rate of expansion (see, e.g., Ref. \cite{Sh}),
allow one to expect that there is no prolonged post-thermal stage of 
hadronic kinetic evolution, 
and therefore observed particle momentum spectra are not strongly influenced by the final-state hadronic rescatterings
(apart from the Coulomb final-state interactions, decays of resonances, etc.).
 Then,   the saturation of the radius parameter with 
charged-particle multiplicity (see Fig. \ref{fig:6}, left, and
Refs. \cite{Atlas,CMS} )
can indicate increase of the particle number
density at momentum  freeze-out  for large values of charged-particle multiplicity. The latter, according to our analysis,
results in the ground-state condensation.  Such a condensation enhances the coherent properties 
of particle emission   and, therefore, leads to decrease of the $\lambda$ parameter. Interestingly enough,   the 
experimental  $\lambda$ (see Fig.  \ref{fig:6}, right, and Refs. \cite{Atlas,CMS})  are rather small and,
in fact,  smaller than in
relativistic heavy ion collisions,  
indicating the
possibility of the formation of condensates in high-multiplicity $p+p$ 
collision events.
This observation is  not conclusive, however,  because the  $\lambda$ parameter absorbs and reflects many 
effects, in particular, 
particle misidentification,
contribution from decay of long-lived resonances, etc. Because of these complications, 
theoretical description and model fitting of the $\lambda$ parameters have  so far received little attention,
especially in comparison with  the  HBT radii. It seems, however, that  in order  to reveal   
ground-state condensate  contribution to particle momentum spectra one needs to discriminate different
contributions to the $\lambda$, and fit the  $\lambda$ parameters for various energies of collisions.

\section{Conclusions}

In the present paper we   study two-boson momentum correlations at fixed particle 
number constraint in a simple analytically solvable model of a small thermal expanding system. 
For specificity, we use parameter values 
that correspond  roughly to the values at the system’s breakup in $p+p$ collisions at the LHC energies.
We show that correlation strength parameter $\lambda$ 
decreases with multiplicity and that the HBT radius parameter tends to a constant   at moderately large momenta  
when  multiplicity increases. Both effects take place also at zero expansion velocity, see Ref. \cite{Akkelin-1},
and are associated with the increase of the ground-state fraction $N_{0}/N$ at fairly large $N$ when
$N$ increases. Furthermore,
we find that the interferometry radius parameter 
at  fixed multiplicity decreases when the flow increases and that the same is valid 
for correlation strength parameter $\lambda$. While the decrease of the interferometry radius parameter 
takes also place  for averaged over
multiplicities inclusive
measurements of emission from thermalized   expanding systems \cite{Sin-2}, 
 the decrease of the $\lambda$ parameter  is  specific for multiplicity-dependent measurements. 
 We argue that the decrease of the $\lambda$ parameter is conditioned by
the increase  of  the ground-state  contribution to the particle momentum spectra when the flow increases, i.e., by the 
increasing values of the  $n_{N}^{0}(\textbf{p})/n_{N}(\textbf{p})$ ratio. We expect the main points of
our analysis, such as $N$ dependencies of particle momentum spectra and correlations,
to hold if relativistic corrections are taken into account  and suggest that  certain
features of the  multiplicity-dependent measurements
of the Bose-Einstein momentum correlations  in high-multiplicity  $p+p$ collision events
at the LHC can be conditioned  by 
the presence of ground-state condensates. 

We do not  discuss here momentum dependencies of the 
 correlation parameters at fixed multiplicity.  For
such an analysis considered  simple nonrelativistic quantum-field  model of the 
quasiequilibrium state cannot be applied, and relativistic extension of the model should be
necessary.  We hope that our  paper
will help stimulate research efforts in this  and related directions. 

  \begin{acknowledgments}
The research was carried out within the NAS of Ukraine Targeted 
Research Program ``Collaboration in advanced international projects on high-energy physics and nuclear physics'',
 Agreement No. 7/2022 between the NAS of Ukraine and  BITP of the NAS of Ukraine. One of the authors
 (Yu.S.) thanks the Institute of Theoretical Physics of the  University of Wroclaw for financial support, 
 as well as professors of this institute K. Redlich and D. Blaschke for useful discussions.
\end{acknowledgments}


\begin{thebibliography}{99}

\bibitem{Hydro-pp-1}  C. Shen, Nucl. Phys. A
\textbf{1005}, 121788 (2021).

\bibitem{Hydro-pp-2} B. Schenke,  Rep. Prog. Phys. \textbf{84}, 082301 (2021).

\bibitem{Atlas}  ATLAS Collaboration, Eur. Phys. J. C \textbf{75}, 466 (2015).

\bibitem{CMS} A.M. Sirunyan \textit{et al.} (CMS Collaboration), Phys. Rev. C \textbf{97}, 064912
(2018);   J. High Energy Phys. 03 (2020) 14. 

\bibitem{PBM} V.M. Shapoval, P. Braun-Munzinger, Iu.A. Karpenko, Yu.M. Sinyukov, Phys. Lett. B \textbf{725}, 139
(2013).

\bibitem{Akkelin-1} M.D. Adzhymambetov, S.V. Akkelin, Yu.M. Sinyukov,  	Phys. Rev. D \textbf{103}, 116012
(2021).  

\bibitem{Zubarev}  D.N. Zubarev, \textit{Nonequilibrium Statistical Thermodynamics} (New York, Plenum Press, 1974);
D. Zubarev, V. Morozov, G. R\"{o}pke, \textit{Statistical Mechanics of
Nonequilibrium Processes. Volume 1: Basic Concepts. Kinetic Theory}
(Akademie Verlag, Berlin, 1996); \textit{Statistical Mechanics of Nonequilibrium
Processes. Volume 2: Relaxation and Hydrodynamic Processes} (Akademie Verlag, Berlin, 1997);
W.T. Grandy, Jr., Phys. Rep. \textbf{62}, 175 (1980). 

\bibitem{Jaynes} E.T. Jaynes, Phys. Rev. \textbf{106}, 620 (1957);
 Phys. Rev. \textbf{108}, 171 (1957).
 
\bibitem{Guth-2} M.H. Namjoo, A.H. Guth, and D.I. Kaiser, Phys.
Rev. D \textbf{98}, 016011 (2018).

\bibitem{Recurr-1} P.T. Landsberg, \textit{Thermodynamics} (Interscience, New York, 1961);
P. Borrmann and G. Franke, J. Chem. Phys. \textbf{98}, 2484 (1993).

\bibitem{Sin-1} M. Gyulassy, S.K. Kauffmann, and L.W.
Wilson, Phys. Rev. C \textbf{20}, 2267 (1979); M.I. Podgoretsky,
Fiz. Elem. Chast. At. Yad. \textbf{20}, 628 (1989) [Sov. J. Part.
Nucl. \textbf{20}, 266 (1989)]; D.H. Boal, C.-K. Gelbke, B.K.
Jennings, Rev. Mod. Phys. \textbf{62}, 553 (1990);  U.A. Wiedemann,
U. Heinz, Phys. Rep. \textbf{319}, 145 (1999); R.M. Weiner, Phys.
Rep. \textbf{327}, 249 (2000); R.M. Weiner, \textit{Introduction to
Bose-Einstein Correlations and Subatomic Interferometry} (Wiley, New
York, 2000); M. Lisa, S. Pratt, R. Soltz,  U. Wiedemann, Annu. Rev.
Nucl. Part. Sci. \textbf{55}, 357 (2005); R. Lednick\'y, Phys. Part.
Nucl. \textbf{40}, 307 (2009); Yu.M. Sinyukov, V.M. Shapoval,
Phys. Rev. D  \textbf{87}, 094024 (2013). 

\bibitem{Sin-2} Yu.M. Sinyukov,  Nucl. Phys. A \textbf{566}, 589c
(1994);  in \textit{Hot Hadronic Matter: Theory and
Experiment} edited by J. Letessier, H.H. Gutbrod, and J. Rafelski
(Plenum, New York, 1995), p. 309; S.V. Akkelin, Yu.M. Sinyukov,
Phys. Lett. B \textbf{356}, 525 (1995); S.V. Akkelin, Yu.M.
Sinyukov, Z. Phys. C \textbf{72},  501 (1996).
 
\bibitem{Sh} T. Kalaydzhyan,  E. Shuryak,  	Phys. Rev. C \textbf{91}, 054913 (2015).
 
\end{thebibliography}
\end{document}